# Frontiers in Pigment Cell and Melanoma Research

*Fabian V. Filipp, [*] Stanca Birlea, Marcus W. Bosenberg, Douglas Brash, Pamela B. Cassidy, Suzie Chen, John August D'Orazio, Mayumi Fujita, Boon-Kee Goh, Meenhard Herlyn, Arup K. Indra, Lionel Larue, Sancy A. Leachman, Caroline Le Poole, Feng Liu-Smith, Prashiela Manga, Lluis Montoliu, David A. Norris, Yiqun Shellman, Keiran S. M. Smalley, Richard A. Spritz, Richard A. Sturm, Susan M. Swetter, Tamara Terzian, Kazumasa Wakamatsu, Jeffrey S. Weber, Neil F. Box [%]*

**Correspondence**

[*] Fabian V. Filipp, University of California Merced, Systems Biology and Cancer Metabolism, Program for Quantitative Systems Biology, 5200 North Lake Road, Merced, United States, CA 95343;

[%] Neil Box, University of Colorado Denver, Department of Dermatology and Epidemiology, United States, CO, Aurora;

Telephone: 858-349-0349 and 303-724-0160; [*] [%] Email: filipp@ucmerced.edu and neil.box@ucdenver.edu

**Summary**

In this perspective, we identify emerging frontiers in clinical and basic research of melanocyte biology and its associated biomedical disciplines. We describe challenges and opportunities in clinical and basic research of normal and diseased melanocytes that impact current approaches to research in melanoma and the dermatological sciences. We focus on four themes: (1) clinical melanoma research, (2) basic melanoma research, (3) clinical dermatology, and (4) basic pigment cell research, with the goal of outlining current highlights, challenges, and frontiers associated with pigmentation and melanocyte biology.

**Significance**

This document encapsulates important advances in melanocyte and melanoma research including emerging frontiers in melanoma immunotherapy, medical and surgical oncology, dermatology, vitiligo, albinism, genomics and systems biology, epidemiology, pigment biophysics and chemistry, and evolution.

## 1. Introduction

Variability and dysfunction in pigment-producing melanocytes lead to considerable consequences for both human and animal physiology. We describe emerging frontiers in clinical and basic research of normal and diseased melanocytes that impact current approaches to research in melanoma and the dermatological sciences. We focus on four themes, organized in corresponding sections: (1) clinical melanoma research, (2) basic melanoma research, (3) clinical dermatology, and (4) basic pigment cell research, with the goal of outlining current highlights, challenges, and frontiers associated with pigmentation and melanocyte biology.

The council of the Pan American Society for Pigment Cell Research has summoned a comprehensive assessment of the current state of progress in melanoma research and dermatological sciences. Intensive discussions among members of the International Federation of Pigment Cell Societies (IFPCS), the International Pigment Cell Conference (IPCC) [1], the Society of Melanoma Research (SMR) [1], the Skin of Color Society (SOCS), and the Melanoma Prevention Working Group (MPWG) nucleated this perspective and culminated in the current status, challenges, and opportunities of respective thematic research areas (Box et al., 2018). The scope of this perspective was to highlight emerging trends in the four thematic areas, rather than aiming for exhaustive completeness, with a few citations limited to those appearing in the past 12 months.

In the field of clinical melanoma, mechanistic understanding based on genomic and immunological data has led to extraordinary recent progress in diagnosing and treating melanoma tumors. In clinical dermatology, loss of normal pigmentation is mediated by inflammatory, immunological, and genetic determinants. Vitiligo management innovates by reversing skin disease through suppression of autoimmune inflammation and promoting melanocyte regeneration. Progress in management of albinism includes providing precise molecular definition of underlying genetic mutations, limiting ultraviolet (UV) light exposure by physical barriers, and in the future support by pharmacologic and genetic intervention. In melasma and hyperpigmentation areas, the frontlines are molecular understanding of pathogenesis and systematic evaluation of which combinations of topical, oral, and procedural therapies are most effective in treating postinflammatory hyperpigmentation. In basic melanoma and dermatology research, current frontiers include understanding of genomic profiles, signaling pathways, and biophysics of melanins to arrive at comprehensive insights into developmental and biomedical properties of pigment cells.

## 2. Results and Discussion

### 2.1 SECTION 1—CLINICAL MELANOMA RESEARCH

*2.1.1 State of the art of immunotherapy of malignant melanoma*

Immunotherapy has moved to center stage in melanoma treatment with the recent success of trials involving blockade of immune checkpoints (Weber et al., 2017). The cytotoxic T lymphocyte–associated antigen 4 (CTLA4, CD152), and the combination of programmed death 1 (PD-1, PDCD1, CD279) and programmed death 1 ligand (PD-L1, PDCD1L1, CD274) comprise immune checkpoints and negative regulators of T cell immune function. The roles of CTLA4 and PD-1 in inhibiting immune responses, including antitumor responses, are largely distinct. CTLA4 can regulate T cell proliferation early in an immune response, primarily in lymph nodes, whereas PD-1 suppresses T cells later in an immune response, primarily in peripheral tissues. Inhibition of these targets has provided successful modes of intervention, leads to increased activation of the immune system, and has led to new immunotherapies for melanoma and other cancers. Ipilimumab, an inhibitor of CTLA4, is approved for the treatment of advanced or unresectable melanoma. Nivolumab and pembrolizumab, both PD-1 inhibitors, are approved to treat patients with advanced or metastatic melanoma and patients with metastatic, refractory non-small cell lung cancer. In addition, the combination of ipilimumab and nivolumab has been approved in patients without B-Raf proto-oncogene, serine/threonine kinase (BRAF) activating mutations and metastatic or unresectable melanoma. Checkpoint regulators on T cells play vital roles in limiting adaptive immune responses and preventing autoimmune and auto-inflammatory reactivity in the normal host (Najjar et al., 2017). In cancer patients, PD-1 expression can be very high on T cells in the tumor microenvironment, and PD-L1, its primary ligand, is variably expressed on tumor cells and antigen-presenting cells within tumors, providing a potent inhibitory influence within the tumor microenvironment. While PD-L1 expression on tumors is often regarded as a negative prognostic factor, it can be associated with a positive outcome for treatment with PD-1/PD-L1 blocking antibodies, and has been used to select patients for this therapy. Therapy regimens with immune checkpoint blocking antibodies are characterized by responses of long duration, while a minority of patients displays atypical responses. In such cases, progression may precede tumor shrinkage, and a pattern of autoimmune side effects is associated with this class of drugs. Even though excellent efficacy and some complete remissions have been seen in a limited number of melanoma patients, some of whom may be regarded as cured of cancer, many malignancies do not show responses of long duration with these agents. Predicting tumor responses to immune checkpoint blockade remains a major challenge and an active field of research fueled by systems biology data.

*2.1.2 Targeted therapy, combination therapy and treatment resistance*

Inhibition of the mitogen-activated protein kinase (MAPK) pathway using BRAF kinase inhibitors or MAP2K (MEK1/2) inhibitors as single agents improves survival of patients with *BRAF*(V600E)–mutant metastatic melanoma compared with standard chemotherapy. Although BRAF-activated melanomas have initially been the most responsive melanoma subset due to targeted therapies, patients ultimately develop resistance. Accordingly, there is now rapid progress in combinatorial therapeutic approaches, especially ones taking advantage of engaging the immune system to fight cancer cells. Concurrent BRAF and MEK inhibition improved clinical outcomes further over single-agent BRAF inhibitors, with decreased toxicities related to paradoxical activation of the MAPK pathway in *BRAF* wild-type cells (Dummer et al., 2018). More than 95% of patients experienced tumor reduction when treated with MEK inhibitors dabrafenib and trametinib in combination. Although 50% of patients showed disease progression after 12 months, a fraction of patients experienced long-term progression-free survival. The future selection of patients for combined BRAF and MEK inhibition, and the rational design of clinical trials for those who fail this combination will be assisted by analysis of the clinical, genomic, biochemical, and pathological features of those experiencing long-term benefit versus those who progress. Early responses to BRAF inhibitors are a major determinant for long term beneficial outcome. The faster the response, the more long-lasting and deep the remission tends to be. Melanoma phenotypic plasticity describes how tumors adapt and evolve to selective pressure of MAPK blockade. Melanoma cells respond to the selective pressure of MAPK inhibition through diverse mechanisms of adaptations including genomic, transcriptomic, and epigenomic alterations. Cellular models established to quantify responses to clinical dosage of MAPK inhibitors have provided system-level insight into aberrant reactivation of signaling cascades in response to drug resistance. Downregulation of intracellular inhibitory dual-specific phosphatases (DUSPs) eliminates off-switches of kinases and reengages downstream MAPK effectors. Increase in receptor tyrosine kinase (RTK) ligands, either through autocrine production from tumor cells, paracrine secretion from stroma, or systemic production, promotes resistance to kinase inhibitors. By remodeling of cell surface RTK receptors, including insulin like growth factor 1 receptor (IGF1R), receptor tyrosine kinase MET proto-oncogene, or platelet derived growth factor receptors (PDGFRA/B), an adaptation-associated signaling network promotes intracellular signaling reactivating MAPK or AKT targets. In addition, cell adhesion signaling cooperates with RTK signaling. There are several pharmacological strategies to overcome this resistance. Ceritinib, an anaplastic lymphoma kinase (ALK) receptor tyrosine kinase inhibitor approved for use in lung cancer, enhances efficacy of trametinib in melanoma cells through non-ALK mechanisms. Combined inhibition of BRAF and cell cycle via cyclin dependent kinase 4/6 (CDK4/6) has the ability to block aberrant cell proliferation. Determining patient-specific therapeutic responses will have a major impact on defining an optimal fit for a growing spectrum of targeted therapeutic treatments and schedules.

*2.1.3 Melanoma heterogeneity in progression and therapy resistance*

Identification of precision medicine profiles and stratification of skin cutaneous, uveal, mucosal, and acral melanoma are a defined goal of the melanoma research community (Filipp, 2017; Hayward et al., 2017). In melanoma patients, tumor cells activate pathways common to proliferating fibroblasts, endothelial cells, or even inflammatory and immune cells. Tumor cell surface receptors and tumor-secreted proteins are associated with mitogenic and proliferative signaling, migration, invasion, and antiapoptosis. In addition, there is a group of molecules that is not associated with the phenotype of differentiated melanocytes in the skin or surrounding fibroblasts or stroma but that represents re-expressed developmental genes. Several developmental pathway genes, such as those associated with neurotrophin signaling, noncanonical WNT, or NOTCH signaling, are commonly expressed by melanocyte precursors, stem cells of the dermis, and by melanoma cells. However, they are absent in epidermal melanocytes. In the future, systems biology approaches including large-scale omics platforms or computational models will provide mechanistic insight into melanoma heterogeneity and therapy resistance (Krepler et al., 2017; Zecena et al., 2018).

*2.1.4 Melanoma prevention and epidemiology*

Melanoma is a cancer with globally increasing rates. Regions most affected are those predominantly populated by individuals of European ancestry and lighter pigmentation. Across all populations, there are differences in melanoma incidence and mortality by gender and age. In general, melanoma incidence correlates with cumulative UV exposure and is therefore a function of age. However, while there is a greater incidence in older males than females of the same age group, such a trend appears to be reversed earlier in life, pointing to genetic–epidemiological associations. Across the continents, the highest incidence is found in Australia (up to 40 per 100,000) followed by Northern America and Europe (>10 per 100,000). Rare cases (<0.5 per 100,000) are observed in Asia and Africa. Efforts comprise primary prevention—above all—in children and adolescents aimed at reducing harmful UV exposure to prevent the initiation of carcinogenesis. Secondary prevention is focused on screening and early detection of disease or precursor lesions. Lastly, tertiary prevention is aimed at recurring or second primary incidences of disease in combination with targeted chemopreventive efforts in high-risk populations. The multidisciplinary group dedicated to melanoma prevention includes patient advocates, melanoma foundation leaders, basic scientists, epidemiologists, bioinformaticians, social psychologists, and clinician-investigators from dermatology, dermatopathology, medical oncology, surgical oncology, and radiation oncology. Going forward, collaborative efforts are focused on incorporating training and education, smartphone app-based image acquisition, big data-driven technologies, and machine-learning into melanoma prevention and early detection efforts.

2.2 SECTION 2—BASIC MELANOMA RESEARCH

*2.2.1 Oncogenic signaling in melanoma*

Frontiers in basic melanoma research cover a spectrum from molecular and cellular pathogenesis of melanoma initiation to malignant disease progression. Topics include skin-UV interactions, melanocyte carcinogenesis, their resistance to DNA damage and carcinogens, molecular regulators of malignant transformation, mechanisms of tumor progression and metastasis, and determinants of sensitivity to treatment regimen. Based on genomic signatures, it is established that UV damage drives genomic instability, mutagenesis, and carcinogenesis. UV damage in particular fuels the malignant transformation of melanocytes into melanoma. Therefore, limiting DNA damage is important to prevent melanoma development and progression. Melanin offers significant protection against DNA damage in the basal layer of the skin and shields against solar UV radiation-induced DNA damage. Photo-protection by sunscreens can be assessed by measuring UV-induced gene expression changes in the skin. The UV response pathways of melanocortin 1 receptor (MC1R) and endothelin 1 (EDN1) signaling demonstrated convergence in melanocyte DNA repair. Specifically, EDN1 could compensate for loss of the MC1R signaling axis by reducing generation of free radicals and photoproducts as well as enhancing DNA repair *in vitro* in human melanocytes. The MC1R and cyclic AMP signaling axis regulates nucleotide excision repair and accelerated clearance of UV damage by cyclic AMP-activated protein kinase A (PRKACA) dependent phosphorylation of ATR kinase and scaffolding by anchoring proteins. Moreover, increased repair by cAMP signaling also extended to platinum-induced DNA damage and possibly to interstrand cross-linking damage.

For the topics relating to melanoma progression and treatment, improvement of therapy response took center stage. In light of immunotherapy having emerged as a frontline strategy for patients with melanoma, many investigators are actively engaged in determining how melanoma cells position themselves to induce and/or maintain immune checkpoint blockade. For example, in the *BRAF*(V600E) animal model, response to immune checkpoint blockade did not correlate with mutational burden or neoantigen load but rather with clonality of tumors. Immunotherapy responders tended to have fewer distinct tumor clones than nonresponders. Furthermore, circulating tumor DNA can be a useful biomarker for monitoring treatment responses, for defining clonal burden, and to predict which drug combinations best serve the patient for minimal side effects but maximal efficacy. In other work, estrogen analogues that signal through an estrogen-sensing G protein coupled receptor found on melanocytes and melanoma cells reduced the ability of cells to induce immune checkpoint blockade by downregulating major histocompatibility complex, class I components (HLA) expression while at the same time increasing tumor antigen expression. Moreover, the intestinal microbiome influences sensitivity to melanoma immune checkpoint inhibition (Gopalakrishnan et al., 2018; Matson et al., 2018; Routy et al., 2018). Ongoing research on deregulated metabolism, compromised immune homeostasis, composition of gut microbiome, and tumor immune infiltrates indicates a role of gut health in the response to immune checkpoint inhibitors.

Transcription factors and epigenetic rewiring play a pivotal role in melanoma progression and therapy resistance. Histone methyltransferases and demethylases cooperate with members of the transcriptional machinery to modulate oncogenic gene expression. Transcriptional and epigenomic master regulators include components of the polycomb repressive complex and members of the jumonji family, which are dysregulated in melanoma patients, and engage in crosstalk between tumor metabolism and chromatin remodeling. Vitamin D metabolites participate in differentiation of keratinocytes, can control melanocyte homeostasis, and influence melanoma progression. Expression of cytochrome P450 family member 26 (CYP26B1) and the nuclear vitamin D receptor (VDR) involved in vitamin D signaling is dysregulated during melanomagenesis. In contrast, expression of nuclear retinoid X receptor alpha (RXRA) in keratinocytes in the tumor microenvironment is frequently lost during melanoma progression. Expression of EDN1, a direct target of nuclear receptor signaling, is enhanced during UVB-induced melanomagenesis. RXRA in combination with activated CDK4 and oncogenic NRAS proto-oncogene induces spontaneous and acute UVB-induced melanomagenesis. The pathway of G protein subunit *Gnaq/Gna11* is frequently altered in UV-induced melanomas in transgenic hepatocyte growth factor (*Hgf*) overexpression mouse models. Interestingly, early UV exposure triggers *Gnaq/Gna11* mutations in the *Braf*(V600E) mouse model of melanoma.

Salt-inducible kinase (SIK) inhibition induces an anti-inflammatory phenotype used to stimulate eumelanin synthesis to protect against sunburn and photoproduct formation. Pheomelanin synthesis could be a contributing factor for melanoma progression as in the case of amelanotic melanomas. Highlighting melanoma immunotherapy for overcoming the immune checkpoint blockade, it is recognized that parental neoantigen-devoid melanoma cell lines could serve as model of immunotherapy resistance, wherein overcoming neoantigen deficiency may protect against tumor growths. In turn, this might point to the importance of melanocytic antigens. Interestingly, *Vdr* knockout mouse models are opiate-seeking. The latter phenotype could be rescued by vitamin D supplementation, suggesting a connection between UV, vitamin D metabolism, and opiate signaling.

*2.2.2 Animal models*

Animal models continue to contribute critically important insights into pigment cell and melanoma biology (Aktary et al., 2018). Classical genetic approaches in a variety of species, including mice and zebrafish, have resulted in the identification of an extraordinary number of genes and alleles that have strong effects on pigmentation phenotypes. Use of pigment cell-specific promoters to drive transgenes, including relevant genes, reporters, or genetic engineering tools like Cre/CreER$^{T2}$ recombinase has resulted in the generation of widely utilized strains that enable a variety of melanoma models and pigment cell alterations. *Braf*(V600E), *Nras*(Q61K), and cyclin-dependent kinase inhibitor *Cdkn2a*$^{-/-}$ murine models have delivered insights into growth-arrested melanocytic nevus development and release from senescence, and requirements for melanoma initiation. DNA damage, the tumor protein TP53, and the cell cycle regulators CDKN2A and/or RB1 mediate tumor suppressor-induced escape from senescence. Neither CDKN2A nor TP53 is required for BRAF-induced growth arrest, but aberrations in those components facilitate melanoma progression. The complete tumor phenotype requires MAPK and mechanistic target of rapamycin kinase (MTOR complex 1 and 2) activation. MTORC1 perturbation and loss of serine/threonine kinase 11 (STK11, LKB1) abrogate BRAF induced growth arrest. In addition, DNA methyltransferase (*Dnmt*) and glutamate metabotropic receptor 1 (*Grm1*) melanoma models have been developed, which provide a unique focus on noncanonical mitogen activated protein kinase pathways, G-protein activation independent of *Braf*/*Nras* genotypes, epigenetics, and metabolic signaling in malignant melanoma. Importantly, these two last melanoma models conserve characteristics of melanocytic origin. Loss of *Dnmt*s suppressed melanoma in a *Braf*(V600E)/*Pten*$^{-/-}$ mouse model and enhanced tumor latency by reducing AKT phosphorylation. Along the same line, loss of transcription factor POU class 3 homeobox 2 (*Brn2)* induces tumor initiation in a *Braf*(V600E)/*Pten*$^{+/-}$ context. Lymphangiogenesis is important in the early stages of melanoma invasion and metastasis. Using mouse models, tumor cell expression of vascular endothelial growth factors (VEGFC/D) promoted lymph vessel formation, which fuels metastasis. The heparin binding factor midkine (MDK), which signals through integrins as a prolymphangiogenic factor, is actively secreted by melanoma cells. MDK accumulates at sites of future metastasis, establishes new lymph vessel formation, and primes the area for arrival of melanoma cells through transactivation of vascular endothelial growth factor receptor (KDR, VEGFR) expression through MTOR signaling and ras homolog family member small GTPases (RAC1 and RHOA). In human patients, MDK disrupts the structure of the subcapsular sinus of the lymph node, which is relevant for blocking tumor cell invasion into the node. Further, injection of human induced pluripotent stem cells into mouse embryos allows the formation of chimeric animals, stimulating studies to understand the human specificities of melanomagenesis. These approaches continue to be refined and utilized to study important questions in pigment cell and melanoma biology. The recent emergence of CRISPR/Cas9-based genetic engineering approaches has changed the approach to and design of new animal models. These tools are being adapted by a large number of laboratories, and enable somatic genetic editing in both *in vitro* and *in vivo* screens with efficiencies that have not been possible previously.

2.3 SECTION 3—CLINICAL DERMATOLOGY

*2.3.1 Clinical research advances on albinism*

Alterations of pigment cells lead to disease pathologies that are typically compatible with life, with the exception of malignancies and life-threatening forms of Hermansky-Pudlak syndrome and Chediak-Higashi syndrome. Significantly, however, skin conditions may result in psychosocial comorbidities. Albinism is a rare genetically inherited condition affecting approximately one in twenty thousand people in most world populations. Nevertheless, albinism is much more common in sub-Saharan Africa. People with albinism experience visual deficits, hypopigmentation, and sun-sensitivity. In less advanced societies, people with albinism may not have access to healthcare specialists, sunscreen, or protective clothes. Patients are prone to sunburns and UV-dependent skin cancers, particularly in regions with a high UV index. If not removed in a timely manner, these skin cancers can eventually metastasize and cause premature death, often before the age of 40. Sadly, patients with albinism in Africa face a humanitarian crisis. This crisis stems from a lack of access to resources and an alarming increase in ritual assassinations of people with albinism (Franklin et al., 2018). The united pigment cell community stands in solidarity and supports patients suffering from albinism, leaving a unique footprint in the field of pigment cell biology.

*2.3.2 Progress in clinical and basic research of the autoimmune disease vitiligo*

Vitiligo is an acquired chronic depigmenting disorder of the skin, with an estimated prevalence of 0.5% of the general population. Genetic predisposition and environmental triggers likely engage multiple aspects of pigment cell biology, including altered immune and inflammatory responses, cell death pathways, metabolic abnormalities, and impaired cell renewal. In the vast majority of cases, vitiligo has an autoimmune basis and is often associated with other autoimmune diseases. Immune cell infiltrates are frequently found in the perilesional margins in actively depigmenting skin. Recent progress in the understanding of immune pathomechanisms opens avenues for innovative diagnosis and treatment strategies.

Identifying biomarkers to assess disease activity or stability of lesions will be important in the clinical management of vitiligo, understanding of disease pathogenesis, and in the design of clinical trials. Besides measuring levels of chemokines in serum, suction blister fluid from lesional skin can serve as an alternative source. Cytokine levels in blister fluid appear to be sensitive and specific for defining active versus stable disease and treatment response. Furthermore, blister fluid allowed quantitative phenotyping of melanocytes and their lineages, providing insights into vitiligo pathogenesis and repigmentation. Phototherapy remains a mainstay treatment for targeted vitiligo lesions with less than 10% of the body surface affected. The combination of topical steroids such as fluticasone propionate along with narrowband UVB is safe and provides superior repigmentation to narrowband UVB alone. For targeted lesions, laser treatment can be particularly effective for repigmenting vitiligo. This era has begun to see an advent of targeted, molecular therapies for vitiligo. Autoimmunity in vitiligo appears to be driven primarily by T cell-derived interferon-gamma (IFNG) and its target chemokines, CXCL9 and CXCL10. Clinical trials in targeting this cytokine pathway using the Janus kinase (JAK) inhibitor, ruxolitinib, have already been initiated. Additional pathways that synergize with IFNG to promote disease maintenance through autoimmune memory in the skin offer further avenues for targeted immunotherapy. Patients treated with anti-PD-1 therapies for melanoma develop vitiligo-like lesions consisting of multiple flecked lesions. Significantly, such lesions are not associated with the Köbner phenomenon, or other autoimmune comorbidities. In contrast, increased CXCL10 levels in serum of patients developing vitiligo-like lesions are associated with skin infiltration of CD8$^+$ T cells expressing C-X-C motif chemokine receptor 3 (CXCR3), elevated levels of IFNG, and production of tumor necrosis factor alpha (TNFA). The mechanism of phenol-induced vitiligo depigmentation involves specifically activated CD8$^+$ T cell responses against pigmented cells, and not against keratinocytes.

Perilesional skin biopsies from Sinclair swine vitiligo lesions respond to DNA vaccination with a melanocyte antigen sufficient to drive autoimmunity and provide a unique opportunity to study the process of repigmentation in animal-derived skin with human-like properties. A new method for studying the cellular and molecular foundation of vitiligo was presented by combining immunostaining with laser capture microdissection to define the number and phenotype of melanocyte stem cell populations in the hair follicle bulge during narrowband UVB treatment of vitiligo patients (Goldstein et al., 2018). This strategy facilitates identification of pathways that govern repigmentation in human vitiligo.

Currently, clinical dermatologists have the opportunity to build on successes of melanoma research, largely through understanding immune checkpoint pathways. By building on existing genomic profiles and by connecting immunological signatures of melanocyte-derived samples with patient and context-specific questions, dermatological subdisciplines have the potential to rapidly translate molecular insights into the clinic.

*2.3.3 Progress in vitiligo genetics*

Genome-wide association studies have discovered approximately 50 genetic loci contributing to vitiligo risk in European-derived Caucasians. For many of these vitiligo susceptibility loci, the relevant causal gene and DNA sequence variants have been identified. Corresponding genes are involved in immune and apoptotic regulation, melanocyte function, and autoimmune targeting (Spritz and Andersen, 2017). Study of population isolates with a high prevalence of vitiligo (by combining association and linkage approaches) provides a unique opportunity to ascertain the genetic architecture of vitiligo in relatively simple, homogeneous populations.

In vitiligo research, quantifying and standardizing clinical outcome remains a challenge. Accurately quantifying vitiligo extent and repigmentation in response to treatment by computer-assisted technologies are important recent initiatives. Validation of disease extent and target lesion repigmentation using automation and standardized digital photography will be indispensable for this. Patients' viewpoints are congruently important determinants in formulating severity and repigmentation scores, as emphasized during the vitiligo working group meeting.

### 2.4 SECTION 4—BASIC PIGMENT CELL RESEARCH

#### *2.4.1 Photochemistry and pathophysiology of pigments*

Topics in photochemistry and pathophysiology of pigments included bioorganic chemistry of eumelanin, photobleaching of synthetic pheomelanin, neuromelanin in neurodegenerative processes, and the elevated redox potential of pheomelanin compared to eumelanin, thus accounting for its higher pro-oxidant activity. Density functional theory based calculations and the effect of functional groups on the rate of cyclization of quinones of dopamine, dopa, and N-methyldopamine provide ab initio insights in chromophore binding energies, stabilization, and dynamics. New directions include the molecular understanding of insect cuticular pigmentation. Similar to the oxidative polymerization of melanogenic precursors, dopamine and cysteinyl dopamine, in insect pigmentation, sclerotinogenic precursors—N-acetyldopamine and N-alanyldopamine—are utilized.

Chemically-excited electrons take on a key role in photoproduct formation after UV irradiation. During UV-exposure, cyclobutane pyrimidine dimers (CPDs) are created by coordinated opening of two double bonds on adjacent pyrimidine bases (thymine or cytosine) to join the bases in a cyclobutane ring, a reaction possible only after an electron has been excited. Extraordinarily high energies during UV exposure are accompanied by electron excitation and drastically altered orbital shapes, allowing reactions that quantum mechanics otherwise forbids. CPDs then lead to the UV signature mutations underlying melanoma and other skin cancers. Surprisingly, melanocytes were discovered to also generate CPDs for hours after UV exposure ended. Such dark CPDs constituted the majority of CPDs in human melanocytes *in vitro* and mouse skin, and were most prominent in skin containing pheomelanin. The mechanism was found to be excitation of an electron by a chemical reaction involving melanin, superoxide, and nitric oxide, rather than by a photon. Chemiexcitation is familiar in fireflies but was unknown in mammals. Evidently, melanin is carcinogenic as well as protective. Chemiexcitation is therefore a new mode of pathogenesis that contributes to the CPD footprint known to result in UV signature mutations in melanoma genomes. The process may also underlie diseases in tissue never exposed to UV: Inflammation generates superoxide and nitric oxide and these three factors colocalize with melanin or neuromelanin in tissues susceptible to macular degeneration, noise-induced deafness, and Parkinson's disease.

Pathophysiological processes involve considerable physicochemical changes of melanin and its derivatives, affecting biological functions of pigments. Melanin granules of the aging human retinal pigment epithelium can be impaired by sublethal photic stress, leading to phototoxicity. Interestingly, such phototoxic effects mediated by the aging pigment granules can be reversed by natural antioxidants. Consistent with this idea, retinal pigment epithelium melanin from younger donors was found to be more photoreactive, showing less phototoxicity. Investigations of the redox properties of insoluble melanin exemplify how electrochemical detection is able to simultaneously record optical and redox-based outputs. This enables researchers to trace the generation and quenching of radicals that play critical roles as mediator molecules to create oxidized or reduced forms of chromophores. Thus, electrochemistry provides a unique insight into the redox-cycling capabilities of melanin impacting the biological redox homeostasis of pigmented cells.

#### *2.4.2 Genetics of pigmentation*

Recent advancements in genomic profiling reveal that homologous genes in humans are not only involved in pigmentation disease but also determine skin and hair color in vertebrates. In addition, the field of pattern formation includes the attractive study of how genetic variation influences gradients of pigment factors. Comparative animal models together with human genetics are utilized to study the evolution of pigmentary systems. Work on coat pattern formation using the African striped mouse shows how dorsal racing stripes are established in rodents, others mammals, and fish. Using the Zebrafish as a model system, it is possible to study the long-distance influence macrophages play in adult pigmentation pattern development.

It is essential to understand the molecular, cellular and genetic foundations of these patterns that are common across species to help understand the evolutionary processes under phenotypic selection. The transcription factor ALX homeobox 3 (*Alx3*) was identified as a primary regulator of melanocyte maturation through the pattern of expression in the embryo via inhibition of the

*Mitf* gene within the melanocyte, thereby suppressing pigmentation and giving the appearance of light colored hair (Mallarino et al., 2016). Notably, chipmunks have independently evolved a similar dorsal light stripe pattern through ALX3 regulation of MITF. The contribution of agouti signaling protein (*Asip*) as a differentially expressed gene in the patterning process is yet to be fully defined. In human populations, where pigmentation is studied as a variable trait, most interest has focused on why lighter populations arose from darker ancestors upon leaving Africa. Recent genomic insights revealed the distribution of skin colors in African and human ancestral populations. The architecture of skin pigmentation showed unprecedented variation across human populations and revealed different evolutionary influences. Similar mechanisms, gradients of pigmentation, and selection pressures applied to Africans and Europeans. Genetic evidence indicated that the light pigmentation variant at melanosome solute carrier *SLC24A5* was introduced into the Eastern African population by gene flow from non-Africans (Crawford et al., 2017; Martin et al., 2017). The rs1426654(L374F) change has a big effect on skin type in Europeans. However, it was unexpected that this allele seems to have gone back into African populations and is also driving lighter skin across these populations. Other notable findings are the involvement of hitherto unreported genes in pigmentation variation. These are *Mfsd12*, confirmed by testing in animal models, that is mouse grizzled coat color, and associated with dark skin; *Ddb1/Tmem138*, increased expression suggests increased DNA repair being associated with pigmentation responses. The identified SNPs still only explain about 30% of the variance seen in understudied African pigmentation, so other genes remain to be discovered. Currently, the UK Biobank, with more than 500,000 individuals, provides the widest range of pigmentation genotype-phenotype interactions characterized so far.

Human genetic information has potential for implementation into the dermatology clinic for the delivery of personalized medicine. Difficulties and challenges of germline genomic analysis of patients at high risk of melanoma have to be carefully taken under consideration. Using cultured clonal human melanocytes as a model system, the effect of transcription factor interferon regulatory factor 4 (IRF4) in pigmentation biology and interferon-gamma response is under investigation. Congenic cultures with homozygous allele variation in an *IRF4* intronic regulatory region differed with respect to expression of chemokines and immune checkpoints, CCL3, CCL5, CTLA4 and PD-L1.

### 3. Conclusion

Shared success and progress of the pigment cell research community in understanding of cellular and immunological homeostasis of pigment cells but also clinical challenges and hurdles in the treatment of melanoma and dermatological disorders continue to drive future research activities. Investigators in the pigment cell research community have diverse but complementary backgrounds, and include chemists, physicists, basic biological scientists, dermatologists, surgeons, oncologists, epidemiologists, bioinformaticians, and biomedical engineers. The cross-disciplinary nature of the field is unique, unified around a passion for understanding the establishment and maintenance of normal pigmentation, and benign or malignant pigmentary diseases. This fosters the generation of new translational research ideas crossing multiple disciplines.

**Declarations**

**Competing interests**

There is no competing financial interest.

**Ethics approval and consent to participate**

Not applicable.

**Availability of preprint publication**

The manuscript was made publically available to the scientific community on the preprint server arXiv at https://arxiv.org/abs/1808.01869.

**Endnote**

[1] Information on participants of the XXIII triennial International Pigment Cell Conference (IPCC) of the International Federation of Pigment Cell Societies (IFPCS) and the XIV annual Society of Melanoma Research (SMR) Conference along with all meeting abstracts is available at https://doi.org/10.1186/s12967-018-1609-1, https://doi.org/10.1111/pcmr.12622, and https://doi.org/10.1111/pcmr.12656.